\documentclass[smallextended]{svjour3} 
\usepackage{amsmath,amssymb,amsfonts,bbm,times}
\usepackage[dvipsnames]{xcolor}
\usepackage[T1]{fontenc}
\usepackage[normalem]{ulem}
\usepackage{cite}
\usepackage[colorlinks=true,bookmarks=false,linkcolor=RoyalBlue,urlcolor=RoyalBlue,citecolor=RoyalBlue,breaklinks]{hyperref}
\usepackage{graphicx}
\usepackage{bm}
\usepackage{physics}
\def \qcd {D_{\text{\tiny QC}}}
\journalname{Quantum Information Processing}
\begin{document}
\title{Decoherence and classicalization of continuous-time quantum walks on  graphs}
\author{
    Gabriele Bressanini 
    \and 
    Claudia Benedetti 
    \and 
    Matteo G. A. Paris
}
\authorrunning{G. Bressanini {\em et al.}}
\institute{
    Gabriele Bressanini \email{g.bressanini@imperial.ac.uk} 
    \at QOLS, Blackett Laboratory, Imperial College London, London SW7 2AZ, United Kingdom
    \and 
    Claudia Benedetti \email{claudia.benedetti@unimi.it} \\
    Matteo G. A. Paris \email{matteo.paris@unimi.it} 
    \at Dipartimento di Fisica  {\em Aldo Pontremoli}, Universit\'a degli Studi di Milano, I-20133 Milano, Italy}
\date{\today }
\maketitle
\begin{abstract}
We address decoherence and classicalization of continuous-time quantum walks (CTQWs) on  graphs. In particular, we investigate three different models of decoherence, and employ the quantum-classical (QC) dynamical distance as a figure of merit to assess whether, and to which extent, decoherence {\em classicalizes} the CTQW, i.e. turns it into the analogue classical process. 
We show that the dynamics arising from intrinsic decoherence, i.e. dephasing in the energy basis, do not fully classicalize the walker and partially 
preserves quantum features.  On the other hand,  dephasing in the position basis, as described by the Haken-Strobl master equation or by the quantum stochastic walk (QSW) model, asymptotically destroys the quantumness of the walker, making it equivalent to a classical random walk. We also investigate the {\em speed} of the classicalization process, and observe a faster convergence of the QC-distance to its asymptotic value for intrinsic decoherence and the QSW models, whereas in the Haken-Strobl scenario, larger values of the decoherence rate induce localization of the walker.
\end{abstract}
\section{Introduction } 
Quantum walks (QWs) are the quantum-mechanical counterpart of classical random walks (RWs) \cite{kempe003,venegas2012,xia20,kadian21,frigerio21}. 
If the walker evolves in discrete temporal steps, we talk about discrete-time QWs, while in continuous-time quantum walks (CTQWs)  time is a real parameter \cite{aharonov93,farhi98}. In this work, we focus on continuous-time quantum walks. Quantum walks on graphs are  used to describe the transport of energy or information across a given structure in several physical, chemical and biological systems \cite{mulken2011,mulken07,razzoli11,maciel20}.  
They also proved to be powerful tools for building quantum algorithms \cite{ambainis2003,portugal18} and provide a universal model of quantum computation \cite{childs09,Wang20,Herrman2022}.
\par
Experimental implementations of quantum walks unavoidably involve the interaction with an external environment, and a question arises on whether, and to which extent, the open quantum system dynamics of a quantum walker is detrimental to its quantum features. In turn, the thorough characterisation of decoherence effects in CTQWs is a key ingredient for their use in quantum technology and, in particular, to envisage strategies that may mitigate or cancel out noise.
Several studies investigated how noise stemming from the  interaction between the system and the environment affects the dynamics of quantum walks \cite{norio05,plenio08,caruso09,schreiber11,benedetti16,tama17,
benedetti19,Kurt20}. Different effects may emerge, such as localization, transition toward a classically distributed walker \cite{schreiber11,benedetti16,benedetti19}, but also an improvement in excitation transport efficiency  \cite{plenio08,caruso09,tama17,Kurt20}. However, if a QW loses all of its quantum features and becomes a classical RW, then any potential quantum advantage related to the quantum  process is lost as well. 
\par
In this paper, we  study if  decoherence can turn a  continuous-time quantum walk over a graph into the corresponding classical random walk. We refer to this phenomenon as  \emph{classicalization}. In order to make our analysis quantitative, we use a fidelity-based measure of non-classicality, referred to as quantum-classical distance, that quantifies the differences between the dynamics of a quantum and a classical walker on a given graph  \cite{gualtieri20}. Recently, this quantity has been also used as a tool to design optimal 
quantum walks \cite{frigerio22}.
We use this  distance  to assess if and to which extent decoherence  makes the quantum evolution more similar to its classical  counterpart, at least asymptotically, with respect to the ideal noiseless case.
\par
In the following, we study three different models of decoherence.
First, we investigate intrinsic decoherence \cite{Milburn91}, i.e. decoherence in the Hamiltonian basis, then we shift our focus to two different mechanisms of decoherence in the position basis. 
In particular, we consider dephasing induced by the Haken-Strobl master equation \cite{haken73} and then we  analyse  the interplay of unitary and irreversible dynamics using the quantum stochastic walk framework \cite{QuantumStochasticWalk}.
We show that intrinsic decoherence is not able to completely classicalize the walker: residual quantumness is preserved, and the quantum walker dynamics departs from the corresponding classical one. We give the analytic expressions for the asymptotic value of the quantum-classical distance over the complete, cycle and star graphs.
On the other hand,  the two models of decoherence in the position basis completely  suppress  the quantum features of the coherent process, nullifying the  quantum-classical distance. 
\par
As we will see in detail, some general properties of the quantum-classical distance do not depend on the decoherence model: it reaches an asymptotic value which does not depend on the decoherence rate and its qualitative behavior is not dependent on the size of the graph, nor the considered topology. On the other hand, we show that speed of the convergence of the quantum-classical distance to its asymptotic value is influenced by the noise parameter with a non-universal behavior. For intrinsic decoherence and for the quantum stochastic walk model, we observe that a larger value of the decoherence parameter leads to a faster convergence, whereas for the Haken-Strobl master equation we see an inversion,
such that after a threshold value of the noise rate the quantum-classical distance decays slower.
\par
The paper is organized as follows. 
In Section \ref{theory}, we establish notation and briefly review 
classical and quantum walks, the master equation approach for noisy 
quantum walks, and the quantum-classical distance. In Section \ref{deco1}. we investigate the effects of intrinsic decoherence on CTWQs, while Section \ref{sec:basis} is devoted to the study of  decoherence in the position basis. In Section \ref{sec:concl} we draw conclusions, and provide some concluding remarks.
\section{Preliminary concepts}\label{theory}
Let us consider a finite, simple, undirected, connected graph $G(V,E)$, where $V$ is the set of vertices, also called nodes, and $E$ the set of edges. 
The cardinality of $V$ sets the dimension of the graph, i.e. the number of nodes $N=\vert V\vert$.
A graph is uniquely identified by its  Laplacian matrix $L=D-A$, where $A$ is the adjacency matrix and $D$ is a diagonal degree matrix \cite{nica2018}. 
In particular,  $A_{jk}=1$ if vertices $j$ and $k$ are connected  (with $j\neq k$), and zero otherwise, and $D_{jj}=d_j$ is the vertex degree of the node, i.e. the number of edges connecting the $j$-th node to the others.
$L$ is a positive semi-definite matrix and completely determines the classical  evolution of a walker  on a graph. The dynamics of a classical walker initially localized on node $j$ may be described by the diagonal density matrix 
\begin{equation}
    \mathcal{E}_C [\rho_j](t)=\sum_{k=1}^{N} p_{kj}(t) \vert k\rangle\!\langle k\vert .
    \label{classical}
\end{equation}
 where 
 $p_{kj}(t)=\langle k\vert e^{-\nu L t} \vert j\rangle$ 
 is the classical transition probability from node $j$ to node $k$, $\nu$ is the transition rate,  $\lbrace\vert k\rangle\rbrace_{k=1}^N$ is the complete orthonormal basis which describes localized states of the walker on  the $N$ nodes of the graph and $\rho_j = \vert j\rangle\!\langle j\vert$ is the initial state. 
 The 
 map $\mathcal{E}_C$ describes a continuous-time random walk (CTRW) on the graph 
 $G$.  On the other hand, the purely coherent quantum evolution of a CTQW is obtained by promoting the Laplacian to the system's Hamiltonian $H =\nu L$ \cite{farhi98} (though this is not the only possible choice for a CTQW generator \cite{frigerio21,wong2016}). Without loss of generality we can set the transition rate to $\nu = 1$ and  $\hbar=1$.
The unitary evolution of the quantum walk is thus given by 
$\ket{\psi(t)}=e^{-iHt}\ket{j}$, where $\ket{j}$ denotes the initial localized state.
\subsection{Noisy quantum walks}
If the walker interacts with an external environment, its evolution is no longer unitary and can be described, within the Markovian approximation, by a master equation in the Lindblad form, i.e.
\begin{align}
    \frac{d\rho(t)}{dt}&=-i\left[L,\rho(t) \right]+\sum_{k}\gamma_k \mathcal{D}\left[ O_k \right]\rho(t)\,.
\end{align}
Here $\mathcal{D}$ is the  Limbladian  superoperator, defined as $\mathcal{D}\left[ A\right] \rho = A\rho A^{\dagger}-\frac{1}{2}\lbrace A^{\dagger}A,\rho\rbrace$ with the brackets $\{A,B\}=AB+BA$ denoting the usual anticommutator, $ O_k $ are the Lindblad operators and $\gamma_k \geq 0$.
The main goal of this work is to investigate if, and to which extent, the presence of environmental noise makes the quantum evolution of a walker on a graph more similar to its classical  counterpart (that is described by Eq. \eqref{classical}), compared to the ideal noiseless scenario. 
In particular, we consider three models of noisy quantum walks whose dynamics is induced by three master equations that describe dephasing in some basis of choice.
\par
At first, we focus on decoherence in the energy basis, corresponding to the following  master equation 
\begin{align}
    \frac{d\rho(t)}{dt}&=
    -i\left[L,\rho(t) \right]-\frac{\gamma}{2} [L,[L,\rho(t)]]\label{intrdec} \, ,
\end{align}
which describes {\it intrinsic decoherence} of a quantum system. It has been 
derived \cite{Milburn91} by assuming that at short timescales a quantum system evolves as a stochastic sequence of random unitary operators. The parameter $\gamma\geq 0$ is referred to as the decoherence rate, and $L$ is the graph Laplacian. From an operational point of view, intrinsic decoherence corresponds to a randomized quadratic perturbation \cite{l2pert}.
\par
We then consider two different models of decoherence in the site (position) 
basis. The first one corresponds to the Haken-Strobl master equation \cite{haken73}, which reads
\begin{align}
    \frac{d\rho (t)}{dt} = &-i\left[L,\rho(t)\right] +\gamma\,\sum_k\mathcal{D}[P_k]\rho(t) \, ,
    \label{lindblad1}
\end{align}
where $\gamma\geq 0$ is the decoherence rate and $P_k = \ketbra{k}$ are the site-projector operators.
It can be shown \cite{haken73} that the Haken-Strobl master equation corresponds 
to a system with a tight-binding Hamiltonian affected by stochastic noise describing thermal Gaussian fluctuations of the on-site energies. 
The decoherence parameter $\gamma$ is related  to the variance of the thermal fluctuations.
The Haken-Strobl master equation finds applications in the description of a single excitation 
transport, e.g. in an $N$-site chain subject to dephasing induced by the interaction 
with a fluctuating environment  \cite{Rebentrost09}.
\par
A second model of decoherence in the site basis is provided by the quantum stochastic  walk (QSW) approach \cite{QuantumStochasticWalk}. 
The QSW formalism  axiomatically generalizes a classical random process to a quantum stochastic process, where the topology of the underlying graph imposes constraints on the dynamical evolution of the walker.
It is described by the following master equation \cite{Caruso}
\begin{equation}
    \frac{d\rho(t)}{dt} = -(1-p)\,i\,[L,\rho(t)]+p\sum_{kj}
    \mathcal{D}[P_{kj}]\rho(t) \, .
    \label{qsw}
\end{equation}
Here $p\in [0,1]$ quantifies the interplay  between unitary ($p=0$) and irreversible ($p=1$) dynamics and the $P_{kj}=L_{kj}\ketbra{k}{j}$ are the jump operators, with $L_{kj}=\bra{k}L\ket{j}$.
This model generalizes both the QW and the RW on the graph by continuously interpolating between the two dynamics with a single real mixing parameter $p$. 
This model has been employed  to investigate the role of environmental noise in assisting the transport of energy or information across a network, such as a light-harvesting complex in a photosynthetic process  \cite{Caruso,caruso09,tama17,Kurt20}.
\subsection{The quantum-classical distance}
In order to compare the dynamics of different quantum walks models, and 
to quantify the differences between quantum and classical evolutions, we use a recently introduced dynamical distance  based on fidelity \cite{gualtieri20}. Given two dynamical  maps, describing a classical $\mathcal{E}_C[\rho_{cl}]$ and a quantum  $\mathcal{E}_Q[\rho_{cl}]$ evolution, the QC-distance $\qcd(t)$ between them  is defined as \cite{gualtieri20}:
\begin{equation}
    \qcd(t) \doteq 1-\underset{\rho_{cl}}{\text{min}}\, \mathcal{F}\left(\mathcal{E}_C[\rho_{cl}](t),\mathcal{E}_{Q}[\rho_{cl}](t)\right)\, ,\label{qcd}
\end{equation}
where $\rho_{cl}$ is an initial classical state of the walker, i.e. a statistical operator diagonal in the node basis, 
and $\mathcal{F}(\mathcal{E}_C[\rho_{cl}],\mathcal{E}_Q[\rho_{cl}])$ is the quantum fidelity between the classically and quantum evolved states, i.e.  $\mathcal{F}(\rho,\sigma)=\left[\Tr \sqrt{\sqrt{\rho}\sigma\sqrt{\rho}}\right]^2$. 
The minimization  
in Eq. \eqref{qcd} is achieved, for every graph, by a localized  state $\rho_j$ \cite{gualtieri20}, hence we shall focus on such initial states only. 
An interesting feature of this  measure of non-classicality is that, for noiseless unitary quantum dynamics, it has an horizontal asymptote which depends only on the total number of nodes $N$, i.e. $\lim_{t\rightarrow\infty}\qcd(t)=1-\frac{1}{N}$.
In the next Section, we use the QC-distance as a figure of merit to 
assess  the robustness of CTQW against decoherence. In particular, we compute the QC-distance between the  classically evolved  state $\mathcal{E}_C[\rho_j](t)$ defined in Eq. \eqref{classical}, and the quantum state described by a CPTP-map  modeling decoherence in some basis of choice. 
\section{Decoherence in the energy basis}
\label{deco1}
Let us consider a quantum walker subject to the intrinsic decoherence process described 
by the master equation \eqref{intrdec}.
The solution for an initial localized state $\rho_j=\ketbra{j}{j}$, is given by
\begin{align}
  \mathcal{E}_{L} & [ \rho_j ](t) =
  \sum_{n,p} \langle\lambda_n\vert 
 \rho_j
 \vert\lambda_p\rangle  e^{-i(\lambda_n -\lambda_p)t-\frac{1}{2}\gamma (\lambda_n - \lambda_p)^2 t} \vert \lambda_n\rangle\!\langle\lambda_p \vert \, ,
  \label{milbsol}
\end{align}
where $\lbrace\lambda_n \rbrace_{n=1}^N$ and $\lbrace\vert\lambda_n \rangle\rbrace_{n=1}^N$ are respectively the eigenvalues and eigenvectors of $L$, i.e. $L\vert\lambda_n \rangle = \lambda_n \vert\lambda_n \rangle$. 
Eq. \eqref{milbsol} represents decoherence in the energy basis, and may also be expressed in the following form:
\begin{align}
    \mathcal{E}_L[\rho_j](t)=\int dy \, g(y\vert\, 0,\sigma) \,\,e^{-iL(t+y)} \rho_j e^{iL(t+y)}\, ,
\end{align}
where $g(y\vert \,0,\sigma)$ is a Gaussian probability distribution function with standard deviation $\sigma$ and zero mean value. 
The connection between the two representations is achieved by the correspondence $\sigma^2=\gamma t$. This map can thus be reinterpreted as a coarse graining in time of the ideal quantum evolution.
\par
We now use Eq. \eqref{milbsol} to compute the QC-distance $\qcd(t)$ and analyze 
its behaviour for relevant graph topologies. In the long-time limit, the 
classical transition probability distribution $p_{kj}(t)$
tends to the flat distribution, i.e. for $t\gg 1$ we obtain 
$$\mathcal{E}_C [\rho_{j}](t) \simeq \frac{\mathbb{I}}{N} \quad 
\forall \rho_{j}\,.$$ 
Furthermore, in the same limit,  we see from  Eq. \eqref{milbsol} that, 
because of the exponential damping, the only terms of the sum that survive asymptotically are those that satisfy the condition $\lambda_n=\lambda_p$. We can write the expression for the asymptotic quantum map $\mathcal{E}_L[\rho_j](t_{\infty})\equiv\mathcal{E}_\infty[\rho_j]$ as follows
\begin{equation}
    \mathcal{E}_\infty[\rho_j]=\sum_{n,p} \delta_{\lambda_n \lambda_p}\,\bra{\lambda_n}\rho_j\ket{\lambda_p}\ketbra{\lambda_n}{\lambda_p} \,. \label{intdecrho}
\end{equation}
The asymptotic quantum-classical distance reads
 \begin{align}
    \qcd(t_{\infty})= 1-\min_{\rho_j}\,\left[\left(\Tr \sqrt{\frac{\mathcal{E}_\infty[\rho_j]}{N}} \right)^2\right]. \label{asymptE}
 \end{align}
$\qcd(t_{\infty})$ thus depends on the graph topology but not on the decoherence parameter $\gamma$, which only affects the speed of convergence to its  asymptotic value. 
\par
From the asymptotic expression in Eq. \eqref{intdecrho}, we see that  
decoherence in the energy basis cannot turn the quantum evolution of CTQW into a classical random walk for any graph with a degenerate Laplcian spectrum. Indeed, the QC distance is zero if and only if 
 the stationary quantum state is the maximally mixed state. 
From the expression of $\mathcal{E}_\infty[\rho_j]$ 
it follows that, if the spectrum of $L$ has  eigenvectors corresponding to degenerate eigenvalues $\lambda_n=\lambda_p$ with $\bra{\lambda_n}\rho_j\ket{\lambda_p}\neq 0$ , then the density matrix  has non-zero off-diagonal elements and the QC-distance does not vanish. Indeed, the absence of off-diagonal terms in the quantum density matrix is a necessary, but not sufficient, condition to classicalize the walker. 
The existence of degenerate Laplacian eigenvalues is thus a key element in the classicalization of the QW and it is related to the symmetries of the underlying graph, as we show in Appendix \ref{appSimm}. 
\par
Let us conclude this section with few comments about the speed of convergence of the QC-distance to its asymptotic value. From the expression of the classical transition probability from node $j$ to node $k$
\begin{equation}
    p_{kj}(t)=\langle k\vert e^{ -L t} \vert j\rangle = \sum_{n} e^{-\lambda_n t} \langle k\vert\lambda_n\rangle\!\langle\lambda_n\vert j\rangle
    \label{classicdistrib}
\end{equation}
it follows that the convergence of the classically evolved state $\mathcal{E}_C[\rho_j](t)$ to the maximally mixed state is governed by the smallest non-zero eigenvalue which  is known as the Fiedler value $\lambda_F$. The leading  term in Eq. \eqref{classicdistrib} is $e^{- \lambda_F t}$ and
consequently, the bigger the Fiedler value the faster the convergence of the classical distribution to the uniform one. 
In the quantum case  instead,  the speed of convergence to the  stationary state is determined by the smallest  non-zero value of the  energy gap $(\lambda_n - \lambda_p)^2$, as can be seen from Eq. \eqref{milbsol}.  The parameter $\gamma$  determines the speed of convergence, once  $(\lambda_n - \lambda_p)^2$ is fixed.
For the three classes of graphs we are going to consider in this paper, i.e. complete, cycle and star graphs, the minimum of this gap is achieved by the difference of the two smallest eigenvalues squared, namely when one is zero and the other is the Fiedler value. However, this is not true for a general topology.
\subsection{Complete Graph}
\begin{figure}[h!]
\includegraphics[width=0.99\textwidth]{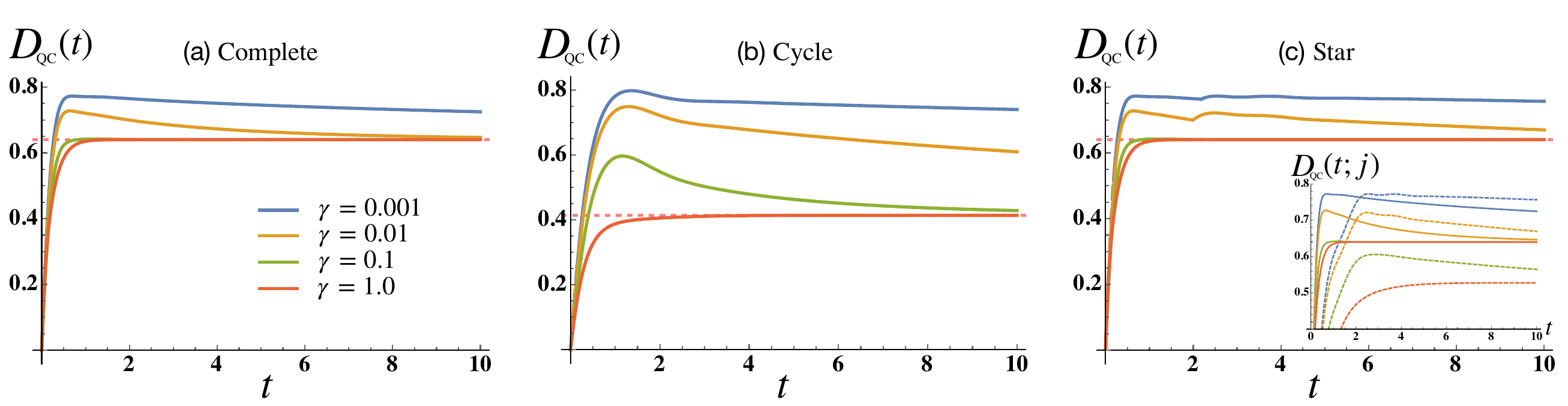}
\caption{Quantum-classical distance computed for different values of the dephasing parameter, for different graphs with $N=5$ nodes: (a) complete graph, (b) cycle graph and (c) star graph. Dashed pink line indicate the asymptotic value $\qcd(t_{\infty})$ in the three considered cases. The inset of figure (c) shows the $\qcd(t;j)$ starting from the two different kind of nodes for the star graph: central (solid lines) and outer (dashed lines).  $t$ is a dimensionless time.  }
\label{plot1}
\end{figure}

The complete graph is the  graph corresponding to maximal connectivity. The Laplacian matrix of a an $N$-node complete graph has the expression $L=N\mathbb{I}-\mathbb{J}$, where $\mathbb{J}$ is the unit matrix, i.e. its elements are all ones, and $\mathbb{I}$ is the identity matrix. In Figure \ref{plot1}(a), we show the QC-distance of noisy (dephased) CTQWs on a complete graph with $N=5$ nodes, for different values of the decoherence rate $\gamma$. All the nodes of a complete graph are equivalent, hence without loss of generality and for future convenience we set the initial state to $\rho_N=\vert N\rangle\!\langle N\vert$.
\par
At the initial time  $\qcd(0)=0$, then the distance grows with time, and eventually reaches an horizontal asymptote, not necessarily monotonically depending on the value of $\gamma$. 
This asymptote does not depend on the decoherence parameter which, in turn, only affects the rate of convergence of $\qcd$ to its asymptotic value. 
In particular, as $\gamma$ gets bigger the distance converges more rapidly. For 
vanishing $\gamma$, the QC-distance for the noiseless case is recovered. 
We observe that intrinsic decoherence indeed brings the quantum walk closer 
to the classical random walk
as the decoherence rate increases.  However, due to the degeneracy of the spectrum, the QW does not become fully classical and the QC-distance remains positive at all times $t>0$. Notice that the qualitative behaviour of the QC-distance does not depend on the number of nodes $N$, i.e. a behaviour similar to that of Fig. \ref{plot1} is observed for every $N$.
\par
Let us now prove that the asymptotic value of the QC-distance does not depend on the rate $\gamma$ but rather it is  a function of $N$ only. The Laplacian 
spectrum of an $N$-node complete graph reads $\lambda_1=\cdots=\lambda_{N-1} = N$ and $\lambda_N=0$. 
The corresponding eigenvectors are 
\begin{equation}
    \vert\lambda_k\rangle=\frac{1}{\sqrt{N}}\sum_{j=1}^N e^{i\frac{2\pi k}{N}j}\vert j\rangle.
    \label{eigenv}
\end{equation}
In order to obtain the stationary quantum-evolved state in Eq.\eqref{intdecrho}, we first need to find the values of $n$ and $p$ for which $\lambda_n=\lambda_p$ holds.
For the complete graph  this condition is satisfied when $n\neq N \wedge p\neq N$ or $n = p = N$. Hence, it follows that the stationary quantum state reads
\begin{equation}
    \rho_{\infty}{=} \frac{1}{N}\vert \lambda_N\rangle\!\langle\lambda_N \vert + \frac{1}{N} \sum_{n,p=1}^{N-1} \vert\lambda_n \rangle\!\langle \lambda_p \vert \, ,
\end{equation}
where we used  $\langle N\vert\lambda_k \rangle=\frac{1}{\sqrt{N}}\,\, \forall k$. We can now compute the asymptotic value of the QC-distance:
\begin{eqnarray}
\qcd(t_{\infty})
=  &&
1-\frac{1}{N^2}\left( 1+\sqrt{N-1} \right)^2.
\label{qcdidc}
\end{eqnarray}
Indeed the asymptotic value does not depend on the decoherence rate $\gamma$, but only on the number of nodes $N$. Notice that for $N\rightarrow\infty$ we retrieve the asymptote of the noiseless case, i.e. $1-\frac{1}{N}$. 
This suggests that, as the size of the complete graph grows, the effect of intrinsic decoherence becomes negligible.
\subsection{Cycle Graph}
The cycle graph is a one-dimensional lattice with periodic boundary  conditions, i.e. it is a regular graph with vertex degrees $d_j=2\quad\forall j$. The Laplacian for a $N$-node cycle graph   is a tridiagonal matrix with off-diagonal elements $L_{j,j+1}=L_{j+1,j}=-1$ $ \forall j=1,\dots,N-1$ and diagonal elements $L_{jj}=2$ $\forall j$. As for the previous case,  and without loss of generality, we choose $\rho_j=\ketbra{N}{N}$ as initial localized state of the walker. 
\par
The behaviour of $\qcd(t)$ is qualitatively similar to the complete graph, see Fig. \ref{plot1}(b), i.e. intrinsic decoherence brings the quantum evolution closer to the classical one compared to the noiseless case.
However, although some quantumness is lost, the degeneracy of the Laplacian's spectrum prevents a full classicalization of the QW.
Depending on the value of $\gamma$, the QC-distance may reach its asymptotic value non-monotonically, i.e. displaying a maximum at shorter times, or monotonically.  We observed that, for a fixed number of nodes, the asymptotic value of $\qcd(t)$ is smaller with respect to the case of the complete graph.
\par
\begin{figure}[t]
\centering
\includegraphics[width=0.5\columnwidth]{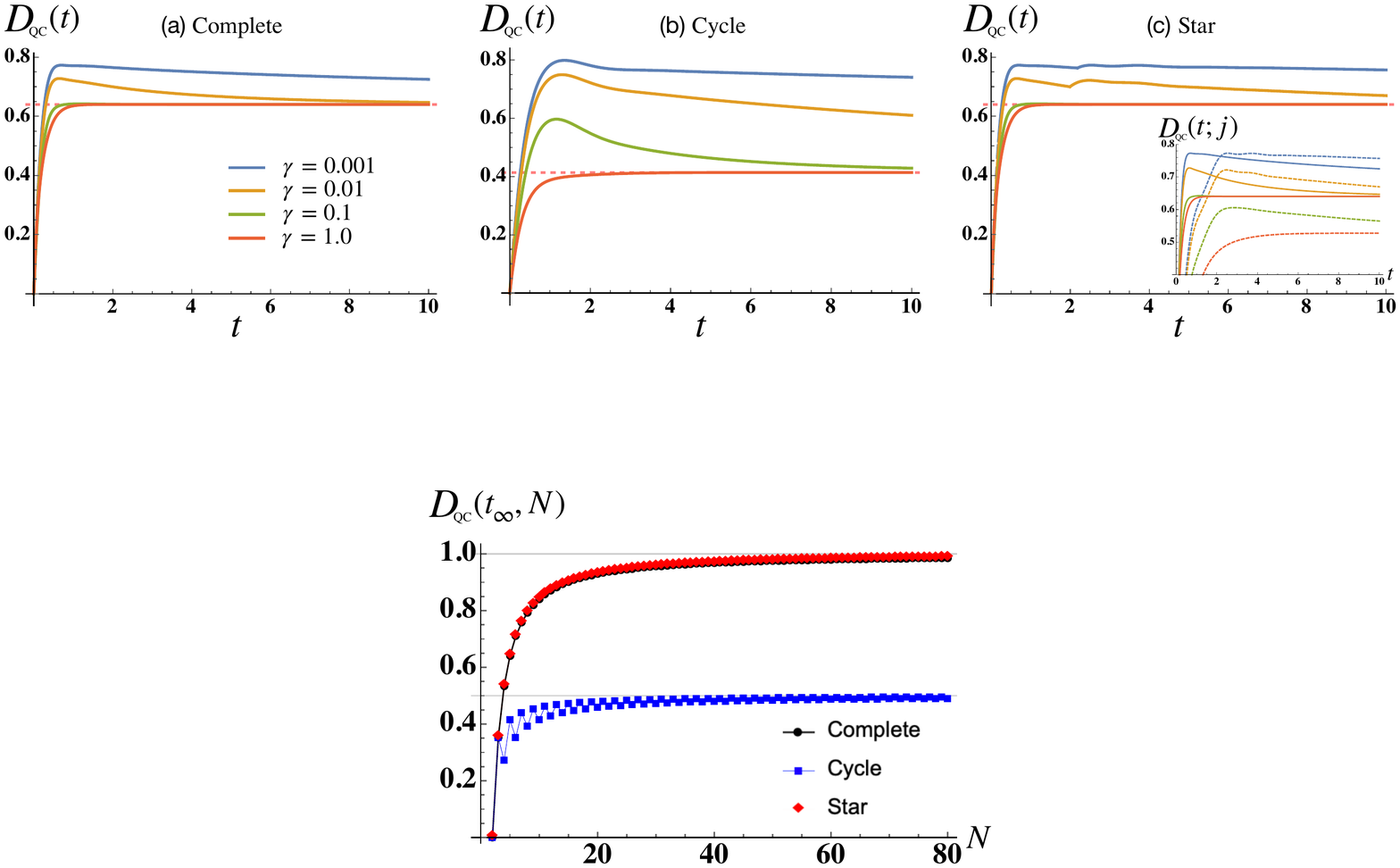}
\caption{Asymptotic value of $\qcd$ in the case of intrinsic decoherence, as a function of the graph size $N$, for the complete (black dots), cycle (blue squares) and star (red rhombus) graph. The values for the complete and star graph are identical.   }
\label{plot1.1}
\end{figure}

To calculate the asymptotic value $\qcd(t_{\infty})$, we need 
$\mathcal{E}_\infty[\rho_j]$  and, in turn, the values 
of $n$ and $p$ that fulfill the condition $\lambda_n=\lambda_p$. 
The eigenvalues of the $N$-cycle graph Laplacian
are $\lambda_n=2\left[1-\cos\left( \frac{2\pi n}{N}\right)\right]$, with $n=1,\cdots ,N$, and the corresponding eigenstates coincide with 
those in Eq. \eqref{eigenv}. In particular, 
$\lambda_n=\lambda_p$ is satisfied when $n=p$ or when $n=N-p$ with $p\neq N$. 
In order to explicitly evaluate  $\qcd(t_{\infty})$, we study the 
odd and the even $N$ cases separately, since when $N$ is even there are $\frac{N}{2}-1$ couples of equal eigevalues, while when $N$ is odd there are $\frac{N-1}{2}$ of them. With this in mind,  we obtain the stationary quantum  state
\begin{equation}
    \mathcal{E}_\infty[\rho_j] = \frac{1}{N} \mathbb{I} + \frac{1}{N} \sum_{p=1}^{N-1}\vert \lambda_{N-p}\rangle\!\langle\lambda_{p}\vert 
\end{equation}
where $p\neq \frac{N}{2}$  if $N$  is even.
After diagonalizing $\mathcal{E}_\infty[\rho_j]$, the asymptote of the QC-distance can be easily computed.
When $N$ is even we obtain
\begin{equation}
    \qcd(t_{\infty}) = 1-\frac{1}{N^2}\left(2+\frac{N-2}{\sqrt{2}} \right)^2 \, ,
\end{equation}
while for odd $N$ we get
\begin{equation}
\qcd(t_{\infty}) = 1-\frac{1}{N^2}\left(1+\frac{N-1}{\sqrt{2}} \right)^2  .
\end{equation}
In both cases, we see that as $N \rightarrow \infty$ the asymptotic value of the QC-distance approaches $\frac{1}{2}$, as opposed to the complete graph case, where the asymptote reaches unity  in the same limit, see Fig. \ref{plot1.1} for a comparison. 
%
%
\subsection{Star Graph}
As opposed to the complete graph and the cycle graph, the nodes of a star graph  are not all equivalent, but are rather grouped into two categories: the central node, which we will refer to as $\vert 1\rangle$, and the remaining $N-1$ external nodes. 
Consequently, the computation of the $\qcd(t)$  involves an optimization over these two distinct classes of   initial  states, consistently with the definition in Eq. \eqref{qcd}.
Fig. \ref{plot1}(c) shows the QC-distance of a quantum walker moving on a star graph. Note that the points of the plot where the function is not differentiable correspond to intersections between the QC-distance of a walker initially localized in the center node and that of a walker whose initial state is an external node.
We first study the system with initial state $\rho_1=\ketbra{1}{1}$ and denote this distance with $\qcd(t,c)$, where $c$ stands for the central node. 
Numerical computation of the distance between the classical and the quantum noisy evolution  shows that $\qcd(t,c)$ is the same as the one obtained for the complete graph.  The same happens in the noiseless case, where  analytic calculations of the quantum and classical transition probabilities  for the two graphs have the same expressions \cite{Xu2009}, as long as the walker is initialized in the central vertex of the star graph.
Moreover, it is possible to show numerically that the QC-distance for the   complete graph  is equivalent to that of a larger class of $N$-nodes graphs, namely those obtained from an $N$-dimensional complete graph by removing edges that are not connected to the central node $\vert 1\rangle$. This holds true for the noiseless scenario \cite{razzoli22} as well as for the dynamics induced by intrinsic decoherence.
\\
If, instead, the initial state is localized on any of the external nodes we obtain a different distance, that we call $\qcd(t, e)$ (see the inset of Fig \ref{plot1}(c)).
In particular,  while at short times the maximum of the $\qcd(t)$ is obtained by starting from the central node, at intermediate times it can be  obtained by starting from one of the external nodes, depending on the value of the parameter $\gamma$. At longer times, the central node always proves to be the optimal initialization for the walker. 
We can indeed compute the asymptotic value of $\qcd(t, e)$ through Eq. \eqref{asymptE}. Given the Laplacian eigenvalues $\lambda_1=\cdots=\lambda_{N-2} = 1$, $\lambda_{N-1}=0$  $\lambda_N=N$, and its respective eigenvectors \cite{Xu2009}:
\begin{equation}
    \vert \lambda_k\rangle= \sqrt{\frac{k}{k+1}}\vert k+2\rangle-\sqrt{\frac{1}{k(k+1)}}\sum_{j=2}^{k+1}\vert j\rangle,
\end{equation}
for $k=1,\cdots,N-2$ and 
\begin{align}
 &   \vert \lambda_{N-1}\rangle=  \frac{1}{\sqrt{N}}\sum_{j=1}^N\vert j\rangle\\
  &  \vert\lambda_N\rangle=\frac{1}{\sqrt{N(N-1)}}\sum_{j=1}^N\vert j\rangle -\sqrt{\frac{N}{N-1}}\vert 1\rangle,
\end{align}
one obtains the following stationary state
\begin{eqnarray}
\mathcal{E}_\infty[\rho_j] =&& \frac{1}{N}\vert\lambda_{N-1}\rangle\!\langle\lambda_{N-1}\vert+\frac{1}{N(N-1)}\vert\lambda_{N}\rangle\!\langle\lambda_{N}\vert + \nonumber \\
+ &&\sum_{j,k=1}^{N-2}\frac{1}{\sqrt{j(j+1)}}\frac{1}{\sqrt{k(k+1)}}\vert\lambda_j\rangle\!\langle\lambda_k\vert.
\end{eqnarray}
The asymptote of the QC-distance, having fixed the initial state to an external node, then reads
\begin{equation}
    \qcd(t_{\infty},e)=  1-\frac{ \left(\sqrt{N(N-2)}+\sqrt{N-1}+1 \right)^2}{N^2(N-1)}
\end{equation}
which is always lower than the one obtained via the central  node as initial state. Hence, the asymptotic distance
\begin{align}
    \qcd(t_{\infty})&=\max\left[\qcd(t_{\infty},c),\qcd(t_{\infty},e)\right]=\qcd(t_{\infty},c)
\end{align}
of the star-graph has the same expression of the asymptotic distance obtained for the complete graph, Eq. \eqref{qcdidc}, as shown in Fig. \ref{plot1.1}.
We can thus conclude that the quantumness of the QW is better preserved asymptotically  if   the walker is initially localized in the  central node of the graph.
\section{Decoherence in the position basis}\label{sec:basis}
Besides energy, the other natural basis to consider for decoherence phenomena is that of the states localized on the graph's nodes $\{\ket{k}\}_{k=1}^N$. 
In the following, we analyze two different  phenomenological models of decoherence in this basis. 
\\
We first consider the  Haken-Strobl  master equation introduced in Eq. \eqref{lindblad1} which, expressed in the node basis, can be recast into  the following form: 
\begin{equation}
    \frac{d\rho (t)}{dt} = -i\left[L,\rho(t)\right] - \gamma (\mathbb{J}-\mathbb{I})\circ \rho(t).
    \label{mastereq_position}
\end{equation}
Here $\mathbb{J}$ is the unit matrix, i.e. a matrix whose entries are all 1, $\mathbb{I}$ is the identity matrix, $\rho$ is the density matrix expressed in the node basis and $\circ$ denotes the Hadamard product of matrices (entry-wise product).
Notice how the  off-diagonal damping terms on the rhs of the equation  do not depend on the distance between the sites.
These dissipative terms cause off-diagonal elements of the density matrix to vanish asymptotically and, at the same time, they also induce a non-trivial dynamics of the diagonal elements since the commutator in the master equation, i.e. the unitary evolution, couples coherences with populations.

\begin{figure}[t]
  \includegraphics[width=\linewidth]{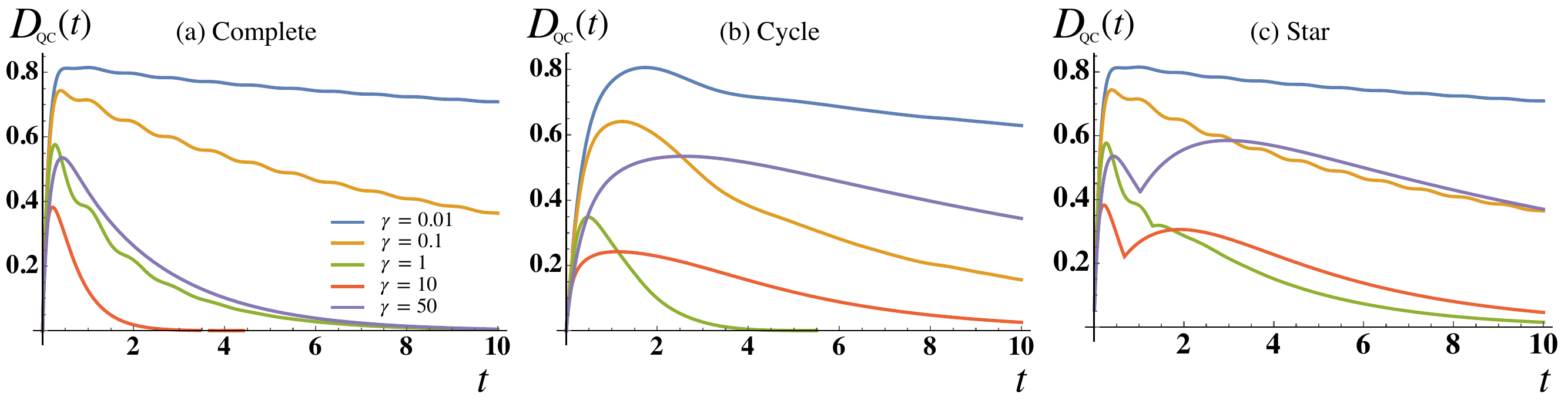}
  \caption{Quantum-classical distance for a walker subject to 
  the Haken-Strobl decoherent dynamics described by Eq. \eqref{mastereq_position} for a (a) complete (b) cycle and (c) star graph  with $N=7$ nodes, computed for different values of the dephasing parameter $\gamma$.}
  \label{figHB}
\end{figure}

We have computed the QC-distance for the complete, cycle and star graphs, and for different values of the decoherence parameter $\gamma$, as in Fig. \ref{figHB}. Some general features of $\qcd(t)$  
do not depend on the kind of graph at study. 
In particular, we find that the quantum-classical distance, after overcoming one or two local maxima,  goes to zero asymptotically. This in turn means that the quantum walk reaches the uniform distribution as a stationary state. 
Hence, contrary to the model of intrinsic decoherence in the energy basis, this map is able to suppress all quantum features of the QW and turn it into the classical random walk asymptotically.
The cuspids in Fig. \ref{figHB}(c) corresponding to the star graph case, arise because of the maximization over classical initial states involved in the definition of  $\qcd(t)$. In particular, for large values of $\gamma$ and at short times, the central node is the one achieving maximal QC-distance, while after the  point where the function is not differentiable, the maximum is obtained by preparing the walker in any of the external nodes. On the other hand, for smaller values of $\gamma$ the  maximization is achieved at all times  by initializing the QW in the central node of the star graph. 

\begin{figure}[t]
  \includegraphics[width=\linewidth]{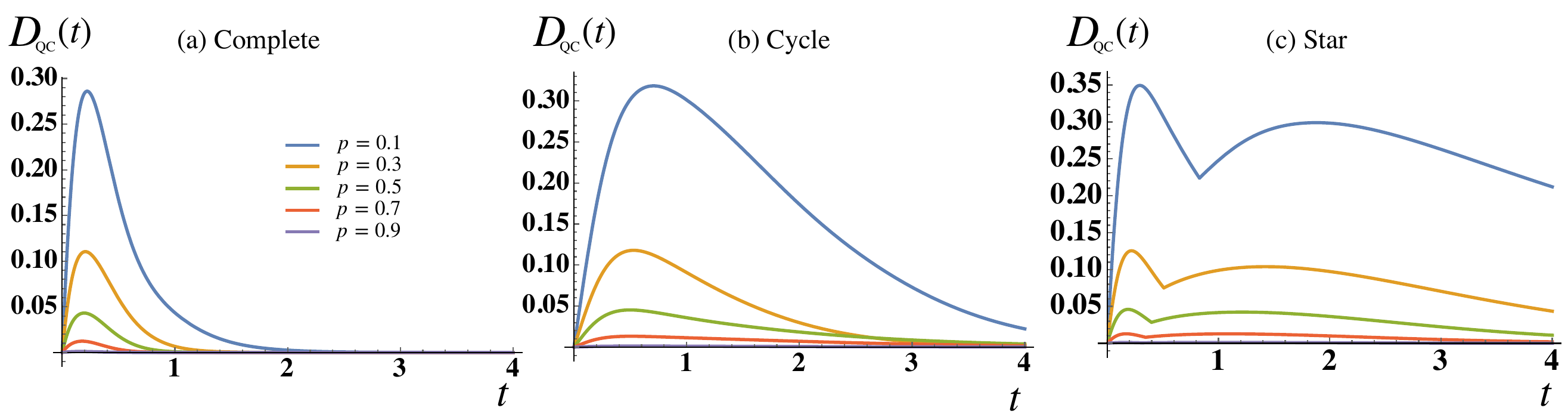}
  \caption{Quantum-classical distance for a walker subject to decoherent dynamics induced by QSW for a (a) complete (b) cycle and (c) star graph  with $N=7$ nodes, computed for different values of the mixing parameter $p$.}
  \label{figSQW}
\end{figure}

Also in this  case, the decoherence parameter $\gamma$  affects the speed of convergence of the QC-distance to its horizontal asymptote.
However, as opposed to the intrinsic decoherence model, it is not true anymore that whenever  $\gamma$ grows so does the speed of convergence of $\qcd$ to its asymptotic value.
In fact, there is a value of $\gamma$, dependent on the dimension and on the topology of the graph, after which there is an 'inversion' and the convergence speed starts decreasing. This fact can be intuitively understood in the asymptotic scenario $\gamma\gg 1$, where the off-diagonal elements of the density matrix experience approximately pure exponential decay, since we can neglect the unitary evolution. 
It is then trivial to show that, within this approximation, the system stays in the initial localized state.
We thus conclude that, as $\gamma$ grows beyond a certain threshold, decoherence effects dominate over the unitary evolution and the system is frozen in the initial state. As a result it takes increasingly more time for the quantum walk to reach the uniform distribution for larger values of $\gamma$.
\\
\\
The second model of decoherence in the nodes basis we consider is the quantum stochastic walk, where the classical and the quantum dynamics are mixed at the master equation level via interpolation through a single real parameter $p$, see Eq. \eqref{qsw}.
We can  solve this master equation numerically and compute the QC-distance for different  values of the mixing parameter $p$ and for the classes of graphs we have considered in this work. 
Our  numerical results show that this kind of noise is able to classicalize the quantum walk asymptotically, regardless of the topology of the underlying graph and for every value of $p>0$. The parameter $p$ only affects the speed of convergence of $\qcd(t)$ to zero, as can be seen in Fig. \ref{figSQW}.
The larger the value of $p$ the faster the QW transitions into a classical random walk, since the classical component in the dynamics has a larger weight in this case, according to Eq. \eqref{qsw}.
We point out that, contrary to the two previous models of decoherence, in this case we find that the QC-distance of a quantum walker initially localized in the central node of a star graph does not coincide with that of a quantum walk on a complete graph. 
The discontinuity in the QC-distance derivative in Fig. \ref{figSQW}(c) signals a change of the initial state that satisfies the maximization condition in the definition (\ref{qcd}). 
In particular, while at short times this condition is achieved by having the initial state localized in the central node, at larger times this is attained by initializing the walker to any of the external nodes. 
As a final remark, we want to highlight the shorter time-scale in Fig. \ref{figSQW} with respect to Fig. \ref{figHB}, indicating a faster classicalization dynamics. The QSW model appears to be the less robust to decoherence for the topologies and choice of parameters considered. 

\section{Conclusions}\label{sec:concl}
Noise is a challenge to practical realisation of  quantum walks. 
Indeed, the presence of noise induces decoherence, and loss of quantum 
features, which are critical to achieve quantum advantages, i.e. to outperform protocols and algorithms based on classical random walks. 
\par
In this work, upon employing a fidelity-based measure of dynamical distance, 
we have addressed  the effects of decoherence on CTQWs. In particular, we have investigated quantitatively the differences between 
the dynamics of a quantum walker in the presence of noise, and that of 
the corresponding classical walker on a given graph. In this way, we have been able to assess if, and to which extent, decoherence makes the quantum walker 
to turn into a classical incoherent random walker.
\par
We have considered three different models of decoherence for a CTQW.
We have first focused on intrinsic decoherence, i.e. decoherence in the energy eigenbasis, and linked  the degeneracy of the Laplacian spectrum to the impossibility for this kind of noise to completely classicalize the quantum walker. In this case, we have analytically characterised the asymptotic 
behaviour of the quantum-classical distance, and discussed how the 
connectivity of the graph 
influences the speed of convergence to its asymptotic value.
We have then shifted our attention towards two models of decoherence 
in the node basis, namely the quantum dynamics arising from the Haken-Strobl master equation, and the quantum stochastic walk model, which interpolates 
between the classical and quantum RW by means of a single real mixing parameter.
In both cases we have shown that all quantum features of the dynamics are suppressed asymptotically, i.e. the quantum walker behaves like a 
classical random  walker in the long-time limit.
\par
Our analysis has shown that, at least for the considered classes of graphs, the qualitative features of the quantum-classical distance are not influenced by the topology of the graph, and by the number of its nodes. In addition, the  
asymptotic value of the quantum-classical distance does not
depend on the relevant noise parameter. Finally, we have found that 
there is no universal behavior of the classicalization speed as a function 
of the decoherence rates. A larger decoherence corresponds to a faster 
convergence of the QC-distance to its asymptotic value for intrinsic decoherence and the QSW models, whereas in the Haken-Strobl scenario, larger values of the decoherence rate induce localization of the walker.
\par
Our results contribute to deepen the knowledge on the effects of noise and decoherence on quantum walks, and pave the way to engineering of 
decoherence, as well as to identifying regimes where its effects  may 
be  mitigated. 

\section*{Acknowledgments}
Work done under the auspices of INdAM-GNFM. G.B. is part of the AppQInfo MSCA ITN which received funding from the EU Horizon 2020 research and innovation programme under the Marie Sklodowska-Curie grant agreement No 956071.

\section*{Appendices}

\appendix
\section{Graph symmetries and degenerate eigenvalues}
\label{appSimm}
In the following we show that the survival of quantum coherences in the asymptotic quantum state Eq. \eqref{intdecrho}, due to the presence of degenerate Laplacian eigenstates, can be traced back to the symmetries of the underlying graph.  
The symmetries of a graph $G(V,E)$ are, in turn, related to the group $\Gamma (G)$ of its automorphisms, i.e. the group of permutations $\sigma: V\rightarrow V$ of the set of vertices $V$ such that    $ \left( \sigma(n),\sigma(p)\right)\in E \Leftrightarrow (n,p) \in E $,   $\forall n,p \in V$, i.e. adjacency is preserved. 
It follows, for example, that for the complete graph every permutation of the nodes identifies a symmetry while, for the star graph, only permutations of the external nodes are automorphisms.  The connection between the symmetries of a graph and the degeneracy of its spectrum is provided by the following theorem \cite{merris}:\\
\\
{\bf {Theorem}}:
Let $G(V,E)$ be a connected graph and $\Gamma (G)$ the group of its automorphisms. If a permutation in $\Gamma (G)$ contains $s$ odd cycles and $t$ even cycles, then the Laplacian matrix will have at most $s+2t$ simple eigenvalues.
\\
\\
It  follows that if a permutation in $\Gamma (G)$ contains a cycle with at least 3  elements, the spectrum of $L$ is degenerate and the QC-distance will not reach zero asymptotically. 
We stress that this is only a sufficient condition for degeneracy. This explains why the QC-distance in Eq. \eqref{asymptE} does not reach zero for the classes of graphs studied in this paper, where it is particularly easy to find symmetries involving three or more nodes. 
On the other hand, we expect to witness a different behavior with other classes of graphs, such as random graphs. In this case, especially for large values of $N$, it is likely that the only symmetries of the graph will be permutations decomposable in cycles of 1 and 2 nodes only. Consequently, in this situation the theorem only tells us that $L$ will have at most $N$ simple eigenvalues, hence it is of no help. 
\bibliographystyle{spphys}
\bibliography{biblioSearch}
\end{document}